\documentclass[twocolumn,showpacs,amsmath,amssymb,showkeys,floatfix,prl]{revtex4}
\usepackage{amsfonts,amsmath}
\usepackage{graphicx}
\usepackage{dcolumn}
\usepackage{bm}

\begin{document}

\title{On detecting CP violation in a single neutrino oscillation channel at very long baselines}

\author{D. C. Latimer$^1$, J. Escamilla$^2$  and D. J. Ernst$^2$}

\affiliation{$^1$School of Liberal Arts and Sciences, Cumberland University, Lebanon,
Tennessee 37087}

\affiliation{$^2$Department of Physics and Astronomy, Vanderbilt University, 
Nashville, Tennessee 37235}

\date{\today}

\begin{abstract}
We propose a way of detecting CP violation in a single neutrino oscillation channel at very long baselines (on the order
of several thousands of kilometers), given precise knowledge of the smallest mass-squared difference. It is shown 
that CP violation can be characterized by a shift in $L/E$ of the peak oscillation in the $\nu_e$--$\nu_\mu$ appearance 
channel, both in vacuum and in matter. In fact, matter effects enhance the shift at a 
fixed energy.  We consider the case in which sub-GeV neutrinos are measured with varying baseline and also the case of a 
fixed baseline. For the varied baseline, accurate knowledge of the absolute neutrino flux would not be necessary; however, 
neutrinos must be distinguishable from antineutrinos.  For the fixed baseline, it is shown that CP violation can be 
distinguished if the mixing angle $\theta_{13}$ were known.
\end{abstract}


\maketitle

\section{Introduction}

The vast majority of neutrino oscillation experiments \cite{solar,kamland,superk,k2k,chooz,minos,mb} can be understood
within a three-neutrino framework \cite{barger_rev,fogli_rev,gg_rev}.
Though oscillation phenomenology is becoming ever more quantitative, there remain unanswered fundamental 
questions regarding the nature of neutrinos.  Oscillation studies will not be able to address some issues; that is, they 
cannot establish an absolute mass scale of neutrinos or whether neutrinos are their own antiparticles.  However, oscillation 
experiments will be able to shed some light on the level of
CP violation in the lepton sector, the relative 
ordering of the mass eigenstates, and the octant of the
$\theta_{23}$ mixing angle.

In the standard theory of neutrino oscillations, flavor interactions are not diagonal with respect to the 
mass eigenstates.  Instead, flavor states are related to mass eigenstates via a unitary mixing matrix $U$ 
which is typically parametrized as in Ref.~\cite{pdg}.  This parametrization contains three angles 
$\theta_{jk}$, which indicate the degree of mixing amongst the eigenstates, as well as three phases.  The two 
Majorana phases cannot be probed in oscillation studies so that the Dirac phase $\delta$ is of sole 
consequence in this realm.  
Finally, 
neutrino oscillations are not sensitive to the absolute mass scale but rather the mass-squared differences, 
$\Delta_{jk}=m_j^2-m_k^2$.  Given three neutrinos, there are two independent mass-squared differences.  A 
current analysis \cite{gg_rev} indicates the one-$\sigma$ values for the mixing angles
\begin{equation}
\theta_{12} = 0.59 \pm 0.02, \quad \vert \theta_{13}  \vert \le 0.09, \quad \theta_{23}= 0.76 \pm 0.08, \label{thjk}
\end{equation} 
and the mass-squared differences
\begin{equation}
\Delta_{21} = 7.9 \pm 0.3  \times 10^{-5} \mathrm{eV}^2, \quad \vert \Delta_{31} \vert = 2.6 \pm 0.2 \times 10^{-3} 
\mathrm{eV}^2.  \label{dmjk}
\end{equation}
At the moment, there is no knowledge of the Dirac phase and only an upper bound exists on the magnitude of 
the mixing angle $\theta_{13}$.   Additionally, the lack of knowledge regarding the mass hierarchy is reflected in our 
ignorance of the algebraic sign of $\Delta_{31}$.

Future experiments will more precisely determine these parameters as well as address our ignorance of the 
remaining unanswered questions.  Herein, we will focus upon the Dirac CP phase $\delta$.  CP violation can 
only be manifest in neutrino oscillations if there are at least three neutrinos with nondegenerate masses 
and nonzero mixing occurs among all three flavors.  Given this, before searching for CP violation, one must 
first concretely establish that $\theta_{13}$ is nonzero. The CHOOZ reactor experiment \cite{chooz} presently
establishes the most stringent bound upon this mixing angle; future reactor experiments will achieve much 
greater sensitivities \cite{dbl_chooz,daya_bay}. Detecting CP violation will be difficult as its effects are 
modulated by the smallness of $\theta_{13}$.  As an added complication, for relatively long baselines 
through the earth, interactions with matter can result in differences between neutrino and 
antineutrino oscillations, even in the absence of intrinsic CP violation.  This ``fake" signal is not a consequence of a 
nontrivial CP phase but is due to the fact that the earth 
is made of matter (and not antimatter).  Finally, ignorance of the mass hierarchy can create difficulties when trying to 
extract intrinsic CP violation from such channels.

To ascertain the level of CP violation in the lepton sector, one might compare neutrino and 
antineutrino appearance channels over the same baseline. This underpins many experimental
proposals; see Ref.~\cite{barger_long} and the references contained therein. 
These experiments require long baselines; however, they are still 
in a regime in which the mass-squared dominance approximation is valid to first order in the ratio 
$\Delta_{21}/\Delta_{31}$. Should $\theta_{13}$ not be too small, one could also, in principle, see  
the effects of CP violation within a single neutrino oscillation channel.
In fact, an index for characterizing the level of CP violation ascertained from a single-channel oscillation
was developed in Ref.~\cite{taka}, as an analog to the usual asymmetry index used when CP conjugate channels are 
operable \cite{Jar}.
Herein, we will consider the effect of CP violation upon a single oscillation channel. 
We begin by examining a neutrino appearance channel in vacuum and then consider neutrinos which 
traverse a constant density mantle; the region of interest will require very long baselines in which terms 
involving $\Delta_{21}$ cannot be linearized.

\section{Vacuum oscillation}
An understanding of the case of vacuum oscillations will guide our efforts. We will use the standard parametrization of 
the PMNS mixing matrix $U$ \cite{pdg}. The probability that an 
$\alpha$-flavor neutrino of energy $E$ will be detected as a $\beta$-flavor neutrino at a baseline $L$ is
\begin{eqnarray}
\mathcal{P}_{\alpha \beta}(L/E) &=&\delta_{\alpha \beta}-4 \sum^3_{\genfrac{}{}{0pt}{}{j >
k}{j,k=1}} \Re (C_{jk}^{\alpha \beta}) \sin^2 (\varphi_{jk})
\nonumber \\ 
&& +2 \sum^3_{\genfrac{}{}{0pt}{}{j > k}{j,k=1}} \Im (C_{jk}^{\alpha \beta}) \sin(2 \varphi_{jk})\,\,,
\end{eqnarray}
with $C_{jk}^{\alpha \beta} := U_{\alpha j} U^*_{\alpha k} U_{\beta k} 
U^*_{\beta
j}$ and $\varphi_{jk} = 1.27 \Delta_{jk} L/E$ where the neutrino mass-squared differences 
$\Delta_{jk} = m_j^2 - m_k^2$ are in eV$^2$ and the ratio $L/E$ is in units of km/GeV.   Recalling that 
the CP phase changes sign when considering antineutrinos, we see that the first sum in the oscillation 
probability consists of CP even terms whereas the last sum consists of CP odd terms provided 
$\alpha \ne \beta$.   In appearance channels $\alpha \ne \beta$, the coefficients of each term in the sum 
are proportional to the Jarlskog invariant $J$ \cite{jarlskog}
\begin{equation} 
\Im (C_{jk}^{\alpha \beta}) = J \sum_{\gamma l} \epsilon_{\alpha\beta \gamma} \epsilon_{jkl}.
\end{equation}
With the standard parametrization one has
\begin{equation}
J = s_{12} c_{12} s_{13} c_{13}^2 s_{23} c_{23} s_\delta\,\,,
\end{equation}
where $s_{jk} = \sin \theta_{jk}$ and so on.
Given that the CP odd terms are modulated by $s_{13}$, the mixing angle $\theta_{13}$ must be appreciable 
to detect CP violation. If one were able to maximize terms linear in $s_{13}$ then CP 
violating effects would be as large as possible. Note that to most easily work with linear in $\theta_{13}$ terms,
we allow for negative $\theta_{13}$ mixing angles as proposed in \cite{angles}. In Refs.~\cite{th13,th13th23}, we noted that such 
terms will be largest whenever the oscillation probability achieves a
local maximum due to the smaller mass-squared difference $\Delta_{21}$; as such, this is a fruitful region in which to 
explore the effects of CP violation.

What then is the effect of the CP phase on a single appearance 
channel?  For the CP even terms, the phase makes an appearance in the form of $\cos \delta$.  Its size can modify the 
amplitude of the oscillation probability to order $\delta^2$. Of more interest are the CP odd terms.  Sizable CP violation 
results in a phase shift of oscillation phases $\varphi_{jk}$; that is, for a given set of mixing angles and mass-squared 
differences, the peak oscillation probabilities will occur at a different value of $L/E$. If the phase shift is significant, 
then perhaps a broadband measurement around a local maximum would measure such a shift; or alternatively, a varying 
baseline experiment could be used to determine the position of the maximum.  This presumes sufficient 
previous knowledge of the mass-squared differences, and this shift in the local maximum cannot independently resolve both 
$\theta_{13}$ and $s_\delta$.

As the two mass-squared differences $\Delta_{21} \approx 8 \times 10^{-5}$ eV$^2$ and $\Delta_{32} 
\approx 2 \times 10^{-3}$ eV$^2$ are well separated, regions exist where oscillations can be approximated 
by a quasi-two-neutrino scenario.  We first examine the region in which oscillations due to the $\Delta_{21}$ 
mass-squared difference is negligible.  To first order, the remaining mass-squared differences 
are degenerate, $\Delta_{31} \approx \Delta_{32}$.  This near equality causes the CP odd terms, given by,
\begin{eqnarray}
\sum^3_{\genfrac{}{}{0pt}{}{j > k}{j,k=1}} \Im (C_{jk}^{\alpha \beta}) \sin(2 \varphi_{jk}) &\approx&
J[\sin(2 \varphi_{31})-\sin(2 \varphi_{32})] \nonumber \\
&\approx& 0.
\end{eqnarray}
to be zero to first order.  

A better region to explore is where the $\Delta_{21}$ driven oscillations are appreciable and the oscillations 
due to the larger mass-squared differences are unresolvable.  In this region, we may approximate
\begin{equation}
\langle \sin^2(\varphi_{3j}) \rangle \approx \frac{1}{2}, \quad \langle \sin(2\varphi_{3j}) \rangle 
\approx  0,
\label{incoh} 
\end{equation}
for $j = 1,2$.  
Here, oscillations $\nu_e \leftrightarrow \nu_\mu$ are the dominant mode, so we will focus upon the 
appearance channel $\mathcal{P}_{e\mu}$.  To first order, this probability is 
\begin{equation}
\mathcal{P}_{e\mu} \approx -2\Re(C^{e \mu}_{31} + C^{e \mu}_{32})-4\Re(C^{e \mu}_{21})\sin^2 (\varphi_{21}) 
-2J\sin(2\varphi_{21}). 
\end{equation}
The constant term due to the unresolvable higher frequency oscillations can be written as
\begin{equation}
-2\Re(C^{e \mu}_{31} + C^{e \mu}_{32}) = 2 s_{13}^2 c_{13}^2 s_{23}^2;
\end{equation}
this term is small as the mixing between mass eigenstates 1 and 3 is small.  The coefficient of the CP even 
term can be written as
\begin{equation}
-4\Re(C^{e \mu}_{21}) = 4[s_{12}^2 c_{12}^2 c_{13}^2(c_{23}^2-s_{13}^2 s_{23}^2) + (c_{12}^2 - s_{12}^2) K],
\end{equation}
where we define $K:= s_{12} c_{12} s_{13} c_{13}^2 s_{23} c_{23} c_\delta$.  

We wish to characterize the effect of CP violation in a single oscillation channel $\mathcal{P}_{e\mu}$.  As such we will 
compare the consequences of maximal CP violation with no CP violation.  First, let us determine the form of the oscillation 
probability whenever CP is conserved. We adopt the convention in Ref.~\cite{angles, gl01} in which we restrict the CP phase 
$\delta \in [0,\pi)$ and allow negative mixing angle $\theta_{13} \in [-\pi/2, \pi/2]$.  For no CP violation $\delta=0$, 
one has $J=0$ and $K=K_\text{max}= s_{12} c_{12} s_{13} c_{13}^2 s_{23} c_{23}$.  Given this, with CP symmetry conserved, 
the $\nu_e$--$\nu_\mu$ oscillation probability is
\begin{equation}
\mathcal{P}_{e\mu} \approx A +B_{CPC} \sin^2 (\varphi_{21})  \label{pemvcpc}
\end{equation}
with $A=2 s_{13}^2 c_{13}^2 s_{23}^2$ and $B_{CPC}=4[s_{12}^2 c_{12}^2 c_{13}^2(c_{23}^2-s_{13}^2 s_{23}^2)+ (c_{12}^2 - 
s_{12}^2) K_\text{max}]$.  Using the best fit mixing angles in Eq.~(\ref{thjk}) for $\theta_{12}$ and $\theta_{23}$ as well 
as the 1-$\sigma$ bound for $\theta_{13}$, one has $A =0.008$ and $B_{CPC} = 0.47$.  As $\theta_{13}$ is small, the 
contribution of the $K_\text{max}$ term to $B_{CPC}$ is on the order of $6\%$.  The first oscillation maximum in this region 
occurs whenever $\varphi_{21} = \pi/2$.  Using the best fit mass-squared difference in Eq.~(\ref{dmjk}), one finds this 
maximum at a baseline to energy ratio of $L/E = 1.57 \times 10^4$ km/GeV.  The maximum value of the oscillation probability 
is 0.48.

When CP violation is maximal $\delta = \pi/2$, one has $J = J_\text{max} =  s_{12} c_{12} s_{13} c_{13}^2 s_{23} c_{23}$ and $K=0$.  Given this, with maximal CP violation, the oscillation probability is
\begin{equation}
\mathcal{P}_{e\mu} \approx A +B \sin^2 (\varphi_{21}) +C\sin(2\varphi_{21}) \label{pemvcpv1}
\end{equation}
with $A$ as above, $B=4s_{12}^2 c_{12}^2 c_{13}^2(c_{23}^2-s_{13}^2 s_{23}^2)$, and $C=-2J_\text{max}$.  With some rearrangement, one may combine the two terms which are dependent upon the oscillation phase to acheive
\begin{equation}
\mathcal{P}_{e\mu} \approx A' + B' \sin^2(\varphi_{21} + \phi)  \label{pemvcpv2}
\end{equation}
with the new terms
\begin{equation}
A' = A+\frac{1}{2}(B - \sqrt{B^2+4C^2}), \qquad B' = \sqrt{B^2 + 4C^2},
\end{equation}
and phase shift
\begin{equation}
\sin (2 \phi) = \frac{2C}{B'}. \label{phishift}
\end{equation}

Comparing with no CP violation, we note two differences:  there is a slight shift in the maximum value of 
$\mathcal{P}_{e \mu}$ and a shift in the value of $L/E$ at which this maximum occurs.  
From the mixing angles in Eq.~(\ref{thjk}), one notes $4C^2 \ll B^2$ so that to a good approximation 
$A' \approx A - C^2/B$. Additionally, there is an adjustment in the depth of the oscillation probability; 
one has $B' \approx B + 2C^2/B$.  Comparing the maximum value of the oscillation probability for the CP 
conserving (CPC) and maximally violating (CPV) cases, we find
\begin{equation}
\mathcal({P}_{e \mu}^{CPC})_\text{max} -\mathcal({P}_{e \mu}^{CPV})_\text{max} \approx 4 \cos(2 \theta_{12}) 
K_\text{max} - \frac{C^2}{B};
\end{equation}
this difference represents a small shift in the maximum value of around $6\%$.

\begin{figure}[ht]
\includegraphics*[width=3in]{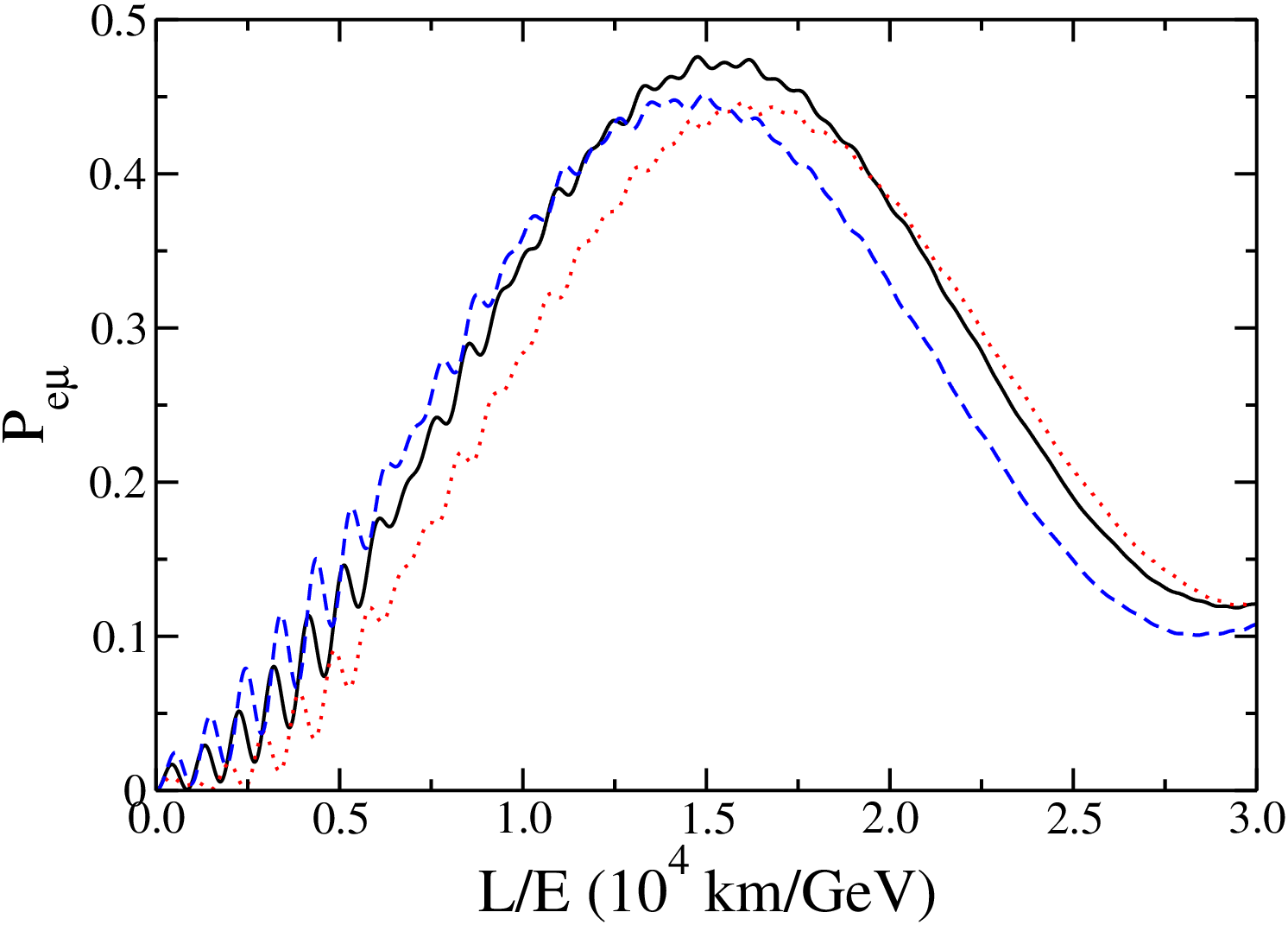}
\caption{[color online] The curves represent $\nu_e$--$\nu_\mu$ oscillation probabilities in vacuum at a long baseline for 
the parameters indicated in the text.  A detector resolution has been included in the curve.  The (black) solid 
curve is the case in which CP is conserved. The (red) dotted curve is the case in which CP is maximally 
violated with $\theta_{13}= 0.09$. The (blue) dashed curve is the case in which CP is maximally violated 
with $\theta_{13}= -0.09$.} 
\label{fig1}
\end{figure}
Perhaps more interesting is the value  of $L/E$ for which these maxima occur.  This is found from the relative phase 
shift $\phi$ between the two sine functions. Using the small angle approximation, one has
\begin{equation}
\phi \approx  \frac{C}{B}\left(1 - \frac{2C^2}{B^2} \right).
\end{equation}
Keeping the first order terms in the mixing angle $\theta_{13}$ and approximating maximal mixing for 
$\theta_{23}$, one has 
\begin{equation}
\phi \approx  -\frac{\theta_{13}}{ \sin(2\theta_{12})}.
\end{equation}
As expected, the shift in phase is proportional to the mixing angle $\theta_{13}$.  Using the usual mixing 
angles from Eq.~(\ref{thjk}), we find a phase shift of $\phi = -5^\circ$. We note that should 
$\theta_{13}=-0.09$ (a negative angle), then the phase shift would be $\phi = +5^\circ$. 

In Fig.~\ref{fig1}, we see that this phase shift results in a shift in the position of the oscillation peak. We compare the $\nu_e$--$\nu_\mu$ 
oscillation probabilities at very long baselines for the 
CPC case and the two CPV cases. We use the mixing 
angles and mass-squared differences from Eqs.~(\ref{thjk}--\ref{dmjk}) and mimic detector resolution so that the rapid oscillations due to the 
larger mass-squared difference are present in the ``wiggles" but are damped out. 
The peak probability for the CPC curve (solid line)  occurs at $L/E=\pi/(2.54 \Delta_{21})
=1.67\times 10^4$ km/GeV as expected.  
Relative to this, the CPV peak probability (dotted line) with positive $\theta_{13}$ leads by an $L/E$ of $0.18\times 10^4$  km/GeV, or 
lags (dashed line) by an equal amount for negative $\theta_{13}$. The height of the peak and the width of the peak 
decrese by a small amount for the CPV cases.

In analyzing such data, it would be key to know the small mass-squared difference, $\Delta_{21}$, with a high 
degree of precision.  Additionally, we have taken for $\theta_{13}$ its 1-$\sigma$ bound; if this mixing angle 
were smaller, the relative shifts in the two probability curves would be proportionally smaller and more 
difficult to detect. Clearly, to make measurements around $L/E \sim 10^4$ km/GeV one would need a {\it very} 
long baseline through the earth's mantle.  For neutrinos of energy 0.5 GeV, for instance, this would occur at a 
baseline of 5000 km.  Matter effects then become quite significant so we next include them in the analysis.

\section{Oscillation in matter}

Neutrinos which travel through sufficiently dense matter undergo coherent forward scattering.  Neutral current 
interactions produce no change in oscillation as all flavors interact equally. Charged current interactions, 
however, result in an effective potential for the electron flavor only  \cite{msw}. In the flavor basis, the 
neutrino evolution equation in matter becomes
\begin{equation}
i \partial_t \nu_f = \left[ \frac{1}{2E} U \mathcal{M} U^\dagger + 
\mathcal{V} \right]\nu_f \label{mswev}
\end{equation}
where the mass-squared matrix is $\mathcal{M} = 
\mathrm {diag} (0, \Delta_{21}, \Delta_{31})$
and the operator $\mathcal{V}$ operates on the electron flavor with a 
magnitude $V= \sqrt{2} G_F N_e$, with $G_F$ the Fermi coupling constant 
and $N_e$ the electron number density. For simplicity we shall assume matter
of constant density. This will avoid the possibility of parametric resonances \cite{parametric}
which could obscure the features we wish to highlight.
For a mantle density of 4.0 g/cm$^3$ \cite{prem,lisi}, 
the matter potential is about $V \sim 1.5 \times 10^{-13}$ eV.
For anti-neutrinos, one needs to change the algebraic sign of this 
potential. For anti-neutrinos, the MSW effect supresses the oscillations and widens the peak so we will confine ourselves 
to the superior
neutrino oscillation channel below.  

For producing results pictured in the graphs, we exactly solve Eq.~(\ref{mswev}). To best understand
the origin of the effects seen,  
it is instructive to examine approximate analytical expressions for the oscillation probabilities.  A constant 
density mantle is a relatively good approximation; however, it limits our baseline
to less than 5000 km \cite{barger8}.  This, in turn, places a limit upon the neutrino energy if we are to reach the peak of 
the $\Delta_{21}$ oscillations. As a result, we  will only consider neutrino energies below 1 GeV. 

For such neutrinos traversing the earth, 
matter affects are most easily understood using the approximations developed in Ref.~\cite{smirnov1}; 
see also Ref.~\cite{th13th23}. With a constant density, one may diagonalize Eq.~(\ref{mswev}) to achieve 
effective mixing angles in matter $\theta_{jk}^m$ as well as modified mass-squared differences $\Delta_{jk}^m$.  For the current scenario, the mass-squared difference of most interest effectively becomes
\begin{equation}
\Delta_{21}^m \approx \Delta_{21} \sqrt{\cos^2 {2\theta_{12}}(1-E/E_R)^2 + 
\sin^2{2\theta_{12} }}, \label{d21m}
\end{equation}
where the resonance energy is
\begin{equation}
E_R = \frac{\Delta_{21}\cos 2\theta_{12}}{2 V c_{13}^2}.  
\end{equation}
The remaining mass-squared differences are modified as well; however, the limit of incoherent 
oscillations used above for vacuum oscillations, Eq.~(\ref{incoh}), is still valid here.  The mixing angle most 
severely affected by the matter is $\theta_{12}$; effectively, one has
\begin{equation}
\sin 2\theta_{12}^m \approx \frac{\sin 2 \theta_{12}}{\sqrt{\cos^2 2\theta_{12} (1-
E/E_R)^2 + \sin^2 2 \theta_{12} }}. \label{th12m}
\end{equation}
The correction also results in a modified mixing angle
\begin{equation}
\theta_{13}^m \approx \theta_{13} + \frac{ \sin{2\theta_{13}}E V}{\Delta_{31}}.
\end{equation}
Finally, the remaining  mixing angle $\theta_{23}$ is unaffected by matter. We note that the only point at 
which a linear power of the mass-squared difference $\Delta_{31}$ enters is in the modification of 
$\theta_{13}$.  This correction for matter effects is quite small so that our results will be independent of 
the (unknown) neutrino hierarchy at the level of a 10\% adjustment to the (already small) $\theta_{13}^m$.  We will 
assume normal hierarchy so that $\Delta_{31} >0$.
\begin{figure}[ht]
\includegraphics*[width=2.95in]{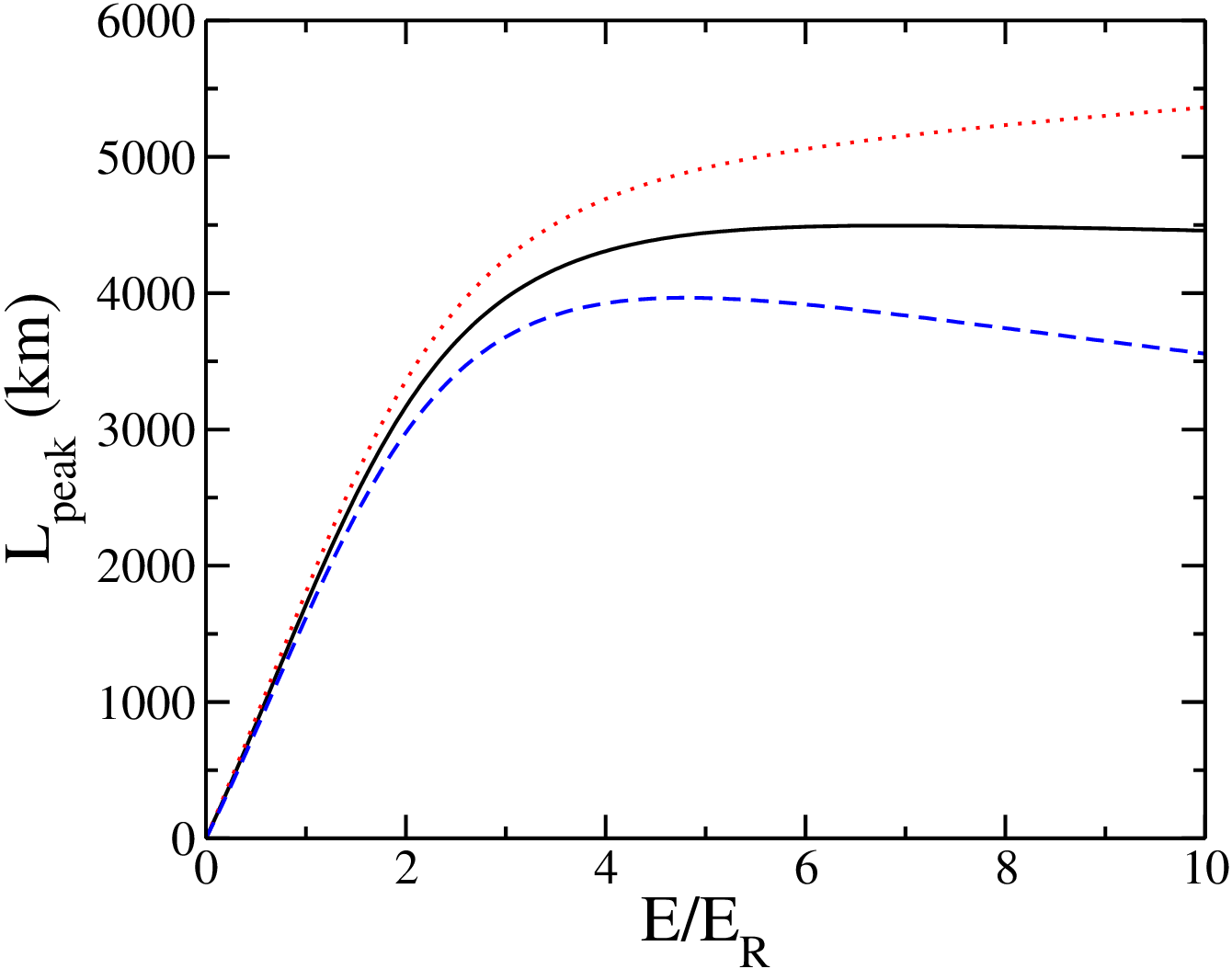}
\caption{[color online] Baseline for the first peak of the $\nu_e$--$\nu_\mu$ appearance probability driven by the effective 
mass-squared difference $\Delta_{21}^m$.  The resonant energy in the mantle is $E_R=0.10$ GeV for the assumed 
neutrino parameters. The (black) solid curve is the case in which CP is conserved. The (red) dotted curve is 
the case in which CP is maximally violated with $\theta_{13}= 0.09$. The (blue) dashed curve is the case in 
which CP is maximally violated with $\theta_{13}= -0.09$.}
\label{fig2}
\end{figure}

In this constant density mantle, the form of the oscillation probabilities developed in 
Eqs.~(\ref{pemvcpc}--\ref{pemvcpv2}) is unchanged; one merely needs to replace all quantities with their 
effective value in matter. From those expressions, we explore the consequences upon the very long baseline 
appearance channel. Given matter effects, the oscillation phase $\varphi_{21}$ now has additional energy 
dependence, thereby modifying the baselines needed to measure the peak of the $\mathcal{P}_{e\mu}^{CPC}$ curve.  
In Fig.~\ref{fig2}, we plot, as a function of energy, the baseline needed to measure the first peak, $L_\text{peak} = 
\pi E /(2.54\, \Delta_{21}^m)$.  In the plot, we express the energy in terms of the resonance energy; for the mixing angles 
and mass-squared differences assumed herein, the resonance energy in the mantle is around 0.10 GeV.
For oscillations in vacuum, the curve would be linear; however, in matter, the baseline 
$L_\text{peak}$ decreases with energy, relative to vacuum oscillations, as the matter term begins to dominate 
the kinetic term in $\Delta_{21}^m$.  
If CP were maximally violated, then the baseline for the peak is shifted according to the phase shift $\phi^m$ 
which also carries an energy dependence by virtue of the modified mixing angles. As with the vacuum case, 
a positive (negative) mixing angle $\theta_{13}$ will result in a longer (shorter) baseline relative to the 
CPC conserved case; this is demonstrated by the dotted (dashed) curve in Fig.~\ref{fig2}.  

\begin{figure}[ht]
\includegraphics*[width=3in]{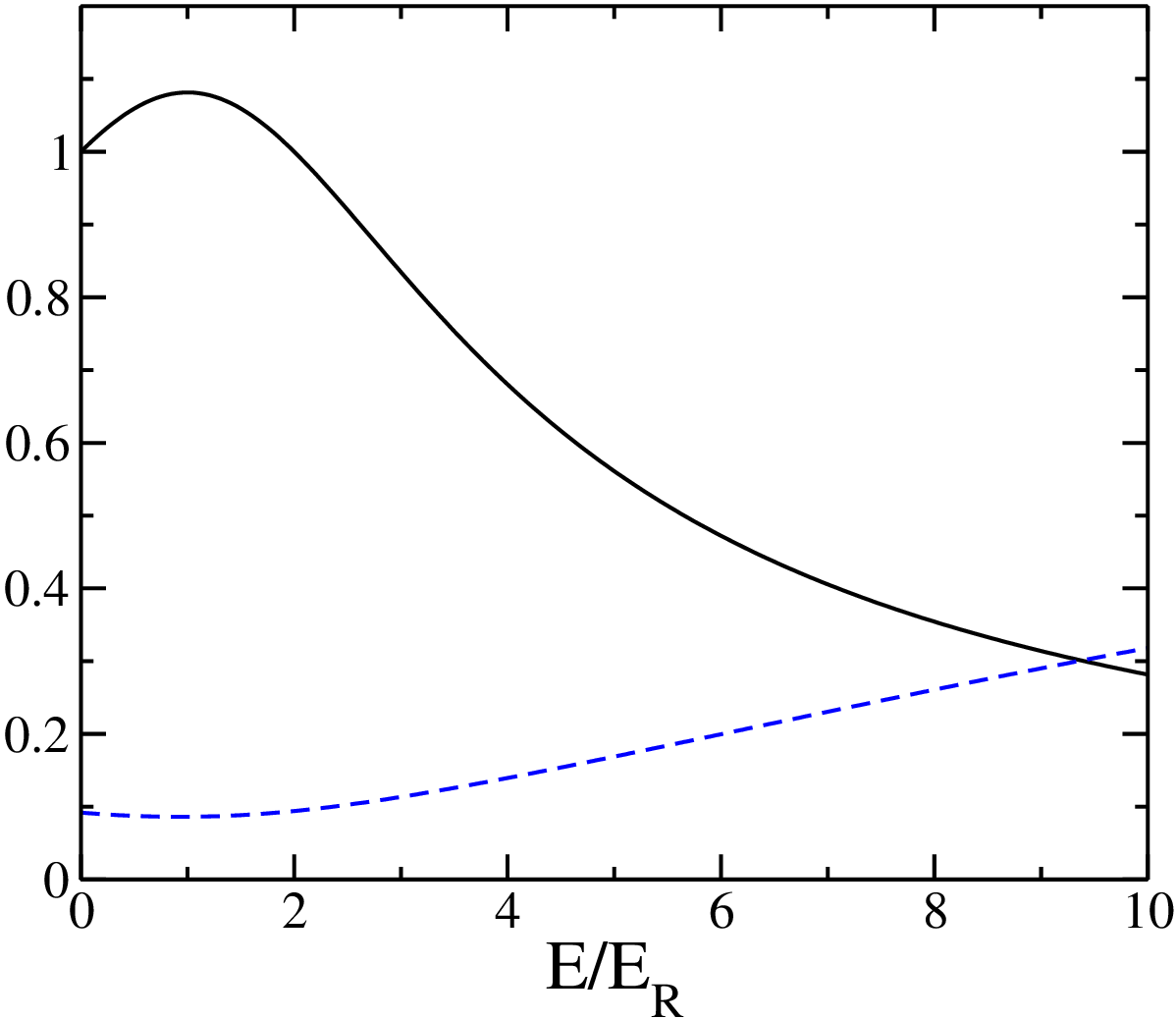}
\caption{[color online] The (black) solid line is the ratio $\sin 2\theta_{12}^m/\sin
2\theta_{12}$.  The (blue) dashed line is the phase shift (expressed in radians) in matter, $\phi^m$, with $\theta_{13}=
-0.09$.} 
\label{fig3}
\end{figure}

The fact that the separation between CPC and CPV peak oscillation probabilities increases with energy is 
attributable to the energy dependence of the phase shift in matter $\phi^m$. Improving on the $5^\circ$ phase 
shift from the vacuum case would aid in resolving the size of the Dirac phase. We recall the definition of 
the phase shift, Eq.~(\ref{phishift}), and substitute the appropriate mixing angles in matter.  The change in 
the mixing angles is most profound for $\theta_{12}$; in fact, we see from Eq.~(\ref{th12m}) that this mixing 
angle vanishes as $1/E$ for large energies.  This is seen in Fig.~\ref{fig3}. 
 
As a result, the term $B^m$ vanishes for large energies resulting in the limit $\phi^m \to \pi/4$.  This 
represents the maximum phase shift between a CPC and maximally CPV peak. The dashed curve in Fig.~\ref{fig3} 
shows how $\phi^m$ increases for the energy range under consideration here. (Note that we use $\theta_{13}=-0.09$ 
so that we may plot a positive phase shift on the same graph.)
The $\theta_{12}$ factor contributes a term linear in energy which also combines with the correction to 
$\theta_{13}$ resulting in a small quadratic contribution.
At $E=0.5$ GeV, the phase shift is about $10^\circ$; for $E=1$ GeV, the shift has risen to $18^\circ$, thus 
enhancing the sensitivity to CP violation.

In addition to the phase shift, the height of the peak value for the oscillation probability will be modified 
by matter effects. Recalling the coefficients $B_{CPC}$ and $B'$, we note that the amplitudes of the 
oscillations are dominated by $\sin^2 2\theta_{12}^m$ in the energy region of interest. As such, the peak 
value for the appearance channel will decrease with energy. With increased energies, matter effects enhance the 
phase shift between the CPC and CPV cases; however, at the same time, the value of the oscillation maxima 
decreases, requiring a more intense beam or larger detector to maintain equal sensitivity.

\section{Varying baseline}

Given the background of the approximate analytical framework above, we shall now show exact numerical results 
which demonstrate the features derived above. We continue with a constant density mantle and use the same 
oscillation parameters as above, but now we solve the propagation equation, Eq.~(\ref{mswev}), sans 
approximation. Our interest is in the position of the peak oscillation in an appearance channel.  Since locating a peak 
can be done without knowing the absolute normalization of the beam (typically a large systematic error in an 
experiment), this approach holds a distinct advantage. To locate a peak, a spread in $L/E$ is required. This can be 
effected by varying the baseline for a fixed energy or measuring a broadband neutrino beam at a fixed baseline.  
\begin{figure}[ht]
\includegraphics*[width=3in]{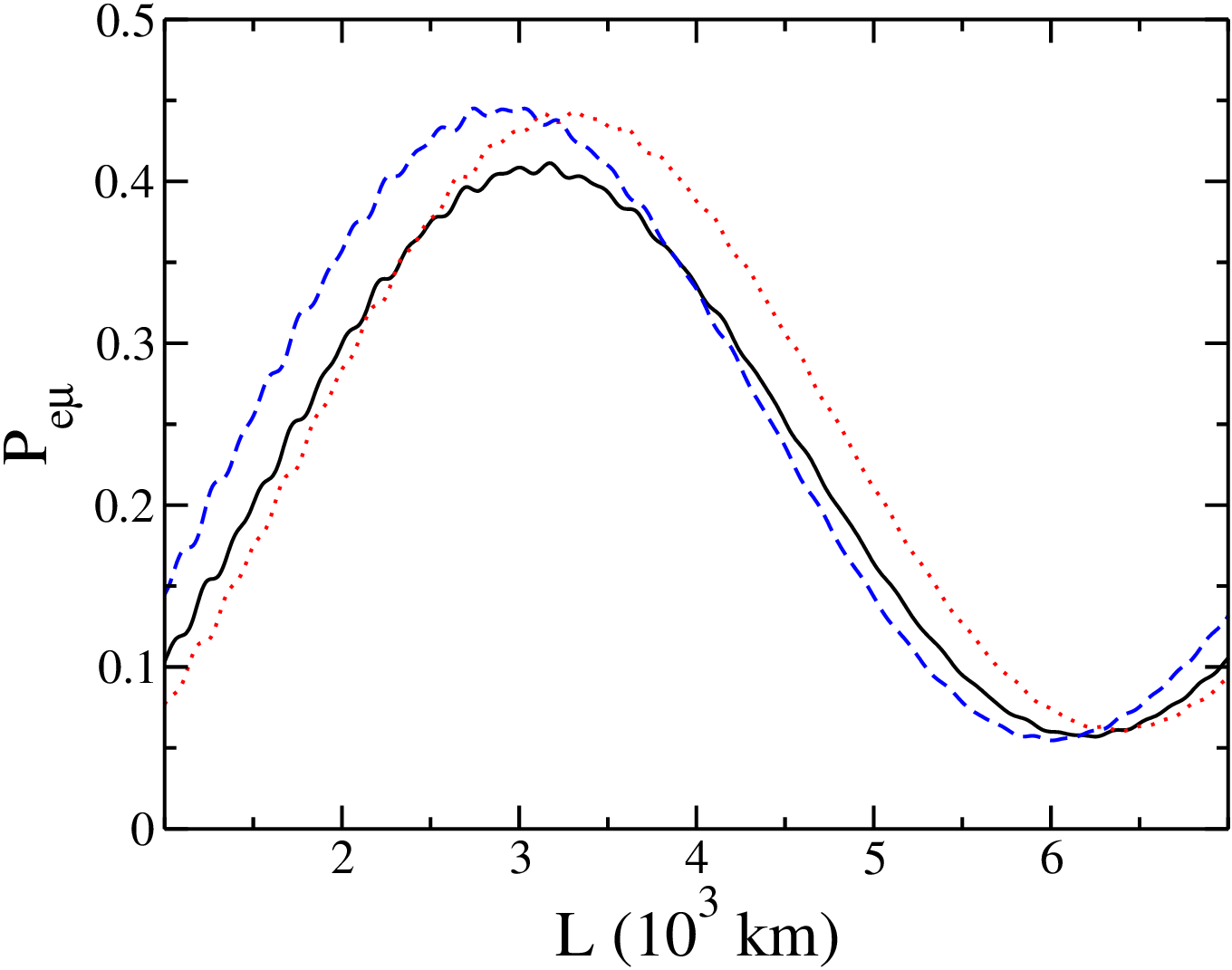}
\caption{[color online] The curves represent $\nu_e$--$\nu_\mu$ oscillation probabilities through a constant density 
mantle at a varying baselines for neutrinos with a narrow energy band of $E_\nu = 0.2$ GeV.  A detector resolution has been included in the 
curves.  CP is conserved for the (black) solid curves. CP is maximally violated with $\theta_{13}=0.09$ for the (red) 
dotted curves. CP is maximally violated with $\theta_{13}=-0.09$ for the (blue) dashed curves.}
\label{fixede1}
\end{figure}

\begin{figure}[ht]
\includegraphics*[width=3in]{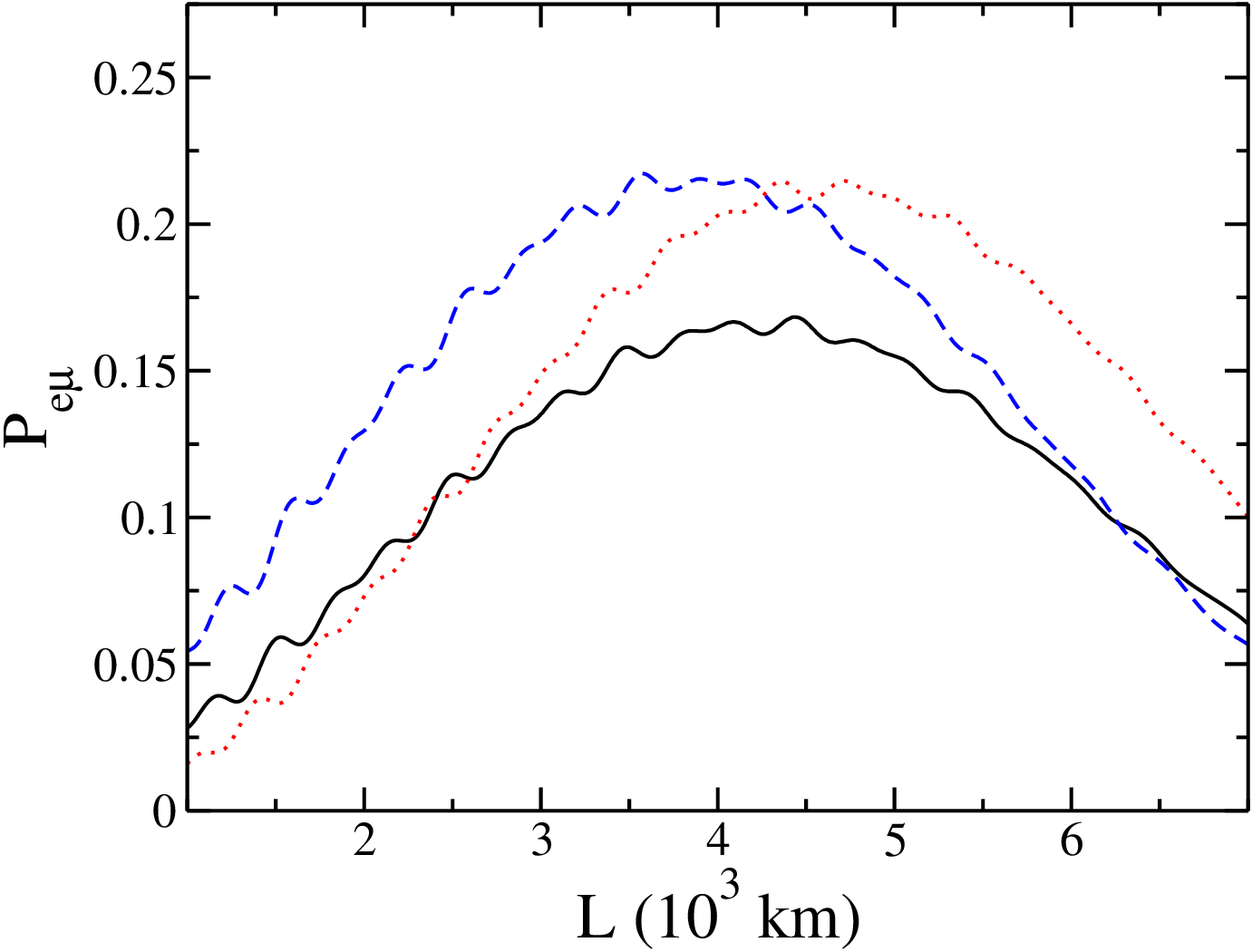}
\caption{[color online] The same as Fig.~\protect\ref{fixede1} except $E_\nu=0.4$ GeV.}
\label{fixede2}
\end{figure}

\begin{figure}[ht]
\includegraphics*[width=3in]{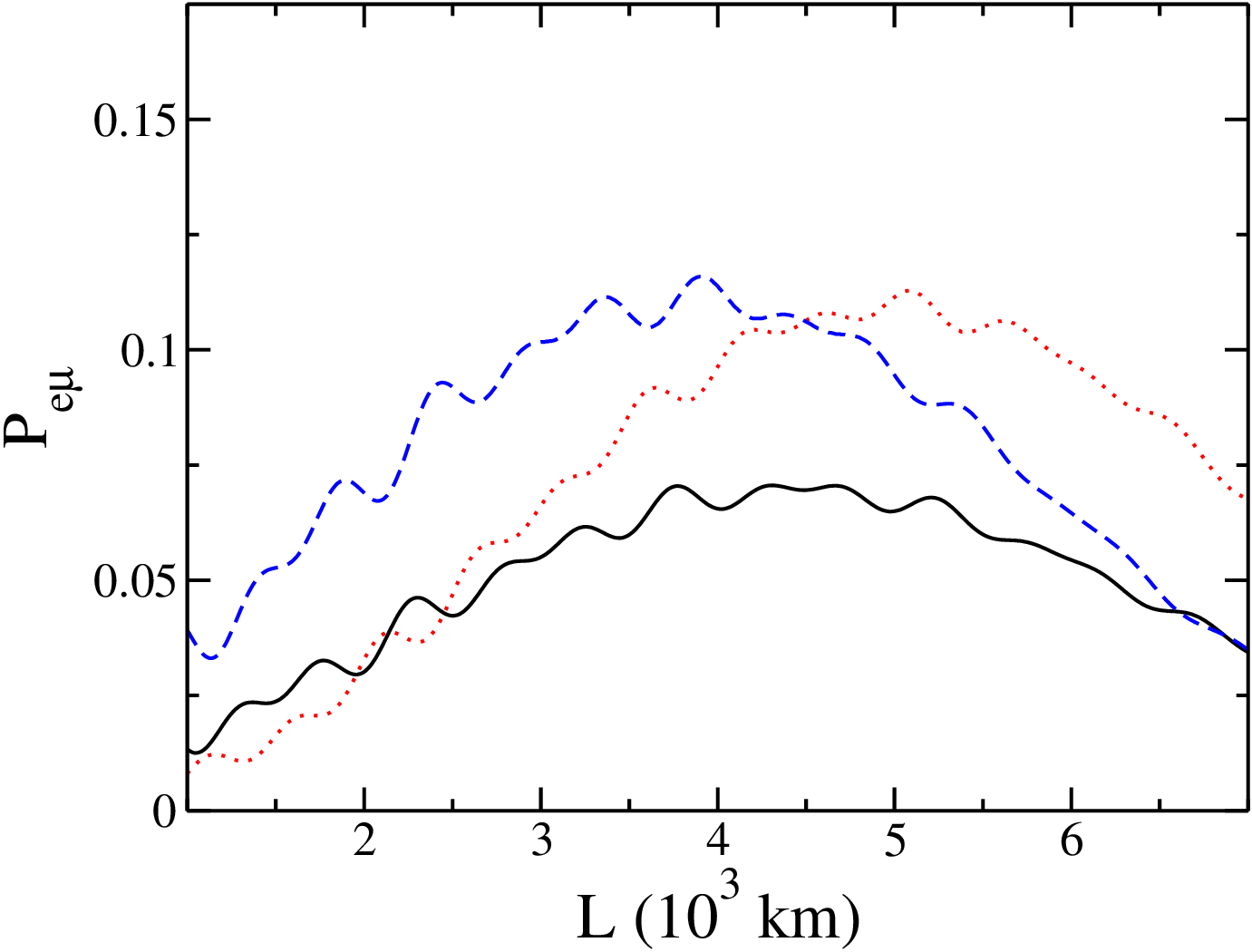}
\caption{[color online] The same as Fig.~\protect\ref{fixede1} except $E_\nu=0.6$ GeV.}
\label{fixede3}
\end{figure}

We first consider a varying baseline for a narrowband neutrino beam. Sub-GeV atmospheric neutrinos would be 
the most likely source for such an experiment. Unfortunately, a water Cherenkov counter such as Super-Kamiokande
\cite{superk} does not have the precision needed to determine the path length of lower energy neutrinos given 
the large scattering angle between the detected lepton and the incident neutrino. Additionally, such detectors 
do not have the ability to resolve neutrino from anti-neutrino which is here needed to achieve a clean signal. 
Future detectors, such as a magnetized iron calorimeter \cite{barger_long}, might have sufficient ability to 
detect the sub-GeV atmospheric neutrinos while identifying whether they are particle or antiparticle. 
Despite the fact that the absolute flux of atmospheric neutrinos is only known to within 15\%, the proposed 
method of detecting CP violation is not sensitive to this as it is the shape of the oscillation probability
curve which is the determining factor. Introducing a finite detector resolution, we plot the appearance 
probability in Figs.~\ref{fixede1}--\ref{fixede3} for three neutrino energies. We include the CPC case and 
the two CPV cases with $\theta_{13}=\pm0.09$.
Notice that the vertical and horizontal axes differ between the three curves. Upon examination, we note that 
the peaks of all curves are roughly located by the approximate baseline 
indicated in Fig.~\ref{fig2}; the separation increases with energy. Note that the height of the peaks decreases
with increasing 
energy, caused by the energy dependence of
$\sin^2(2\theta_{12}^m)$; this decrease in amplitude effectively broadens the peaks.
For energies above 0.5 GeV, the peak of the CPC probability remains 
relatively near 4500 km while the peaks of the CPV curves move farther away from this point.  
For instance, at 0.6 GeV, the peak of the CPV curve with $\theta_{13}=-0.09$ is 3900 km, and the peak for the 
CPV curve with $\theta_{13}=0.09$ is around 5100 km. We note too that there is a relatively significant 
difference in the peak value of oscillation between the CPC and maximally CPV cases. How these features might 
be seen in an atmospheric neutrino experiment is left for future work.

\section{Fixed baseline}

Fixed very long baseline experiments have been previously considered in the literature as a means to uncover 
CP violation, the value of $\theta_{13}$, and discerning the mass hierarchy \cite{barger_long}.  Typically, 
such experiments have baseline and neutrino energies that still permit a dominant mass-squared difference 
approximation to first or second order. As noted previously, we presently consider the other extreme in which the 
oscillations due to the larger mass-squared differences are unresolvable. As a source of neutrinos, we envision an 
accelerator beam stop.  Using an additional detector near the source allows one to have a 
better estimate on the expected flux at a far detector. Also, one can potentially control whether the 
beam is running in a neutrino or antineutrino mode making it inconsequential as to whether the detector can 
distinguish the two. This fact would allow one to make use of existing technologies, such as water Cherenkov 
detectors. In terms of our proposal, present technology would need to be extended to give lower energies 
in order to reach the needed values of $L/E$.  

To consider a fixed baseline detector, we will simulate a broad band neutrino beam with 
energies ranging from 0.1 to 0.4 GeV with a flat spectrum.  As we now are dealing with a range of energies, 
the effective mixing angles and mass-squared differences are no longer constant despite the constant density 
approximation. The phase shift between the CPC and CPV cases will manifest itself differently than in the fixed 
energy case. In Fig.~\ref{fig2}, we plot the baseline for which the oscillation phase $\varphi_{21}^m$ 
achieves the value $\pi/2$ in the constant density mantle, which simplified the analysis for fixed energy.
There is no such simplification for a fixed baseline. 
\begin{figure}[ht]
\includegraphics*[width=3in]{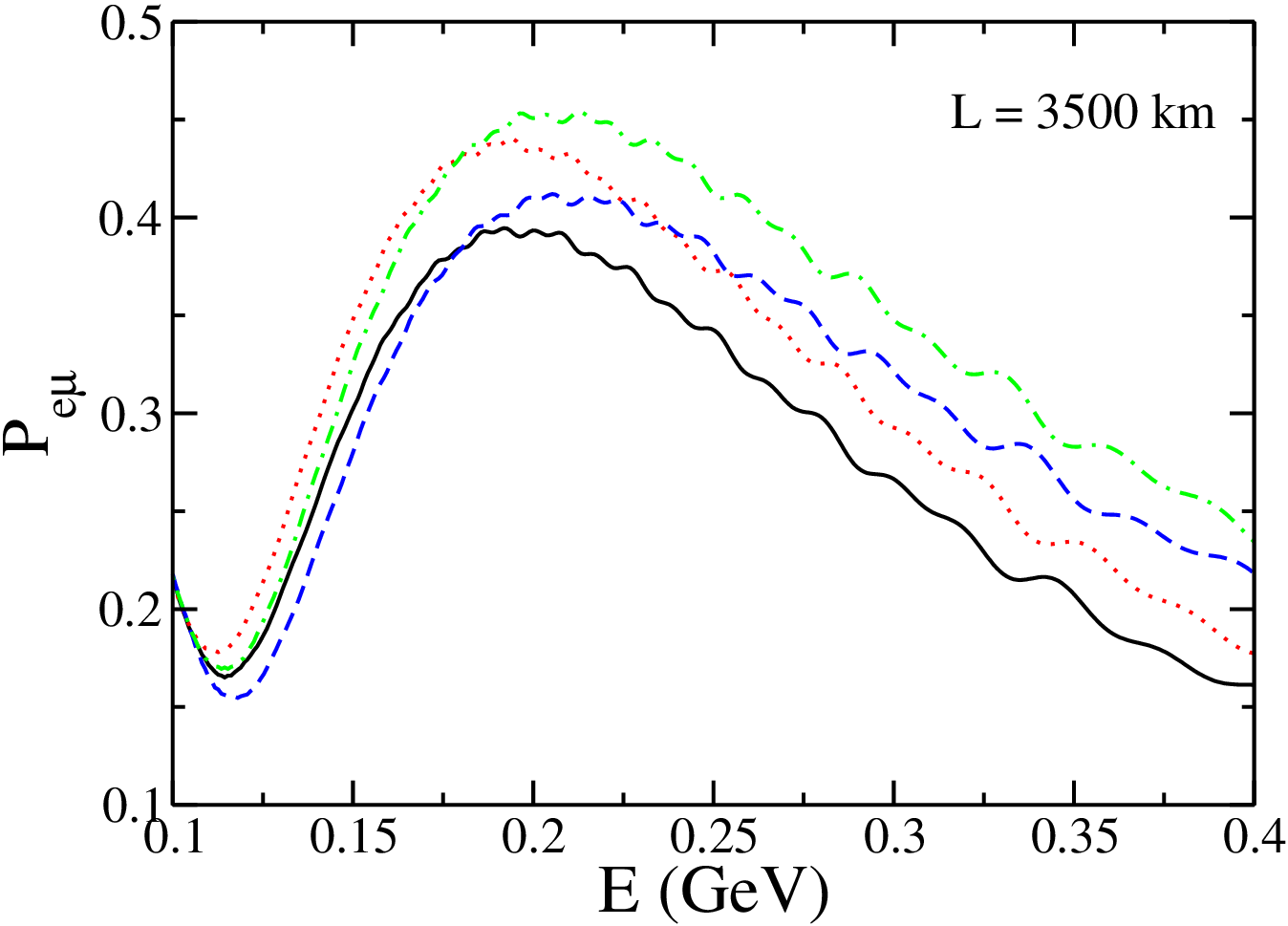}
\caption{[color online] The curves represent $\nu_e$--$\nu_\mu$ oscillation probabilities through a constant density mantle 
at a fixed baseline of 3500 km.  CP is conserved for the 
 (black) solid curve with $\theta_{13}=0.09$ and the (green) dot-dashed curve with $\theta_{13}=-0.09$.
  CP is maximally violated for the (red) dotted curve with $\theta_{13}=0.09$ and the (blue) dashed curve with
   $\theta_{13}=-0.09$.}
\label{fixedl1}
\end{figure}

The maximum value of the appearance channel not only depends upon the oscillatory phase $\varphi_{21}^m$ 
but also upon the energy dependent amplitude of the oscillation. Recall, the energy dependence of the 
amplitude is dominated by $\sin^2(2 \theta_{12}^m)$ which sharply decreases for 
energies greater than twice the resonant energy. On the other hand, for energies below twice the resonant 
energy, oscillation can be enhanced relative to the vacuum. This factor must be included in determining the 
ideal baseline. Incorporating these two factors, we can determine the effect of baseline upon the 
detected neutrino spectrum.  As an example, we examine a baseline around 4500 km.  From Fig.~\ref{fig2},
the oscillatory phases for the maximal CPV case with $\theta_{13}=0.09$ and the CPC case achieve the value of 
$\pi/2$ at 0.35 GeV and 0.62 GeV, respectively. However, as the CPC peak occurs at a higher energy its 
amplitude will be suppressed relative to the CPV case so that the two $\mathcal{P}_{e\mu}$ curves could be 
readily distinguished.  Turning to the CPV case with $\theta_{13}=-0.09$, we note that at this baseline the 
oscillatory phase $\varphi_{21}$ is greater than $\pi/2$ for all energies under consideration. As a result, 
the oscillation probability is further suppressed relative to the other two cases.
\begin{figure}[ht]
\includegraphics*[width=3in]{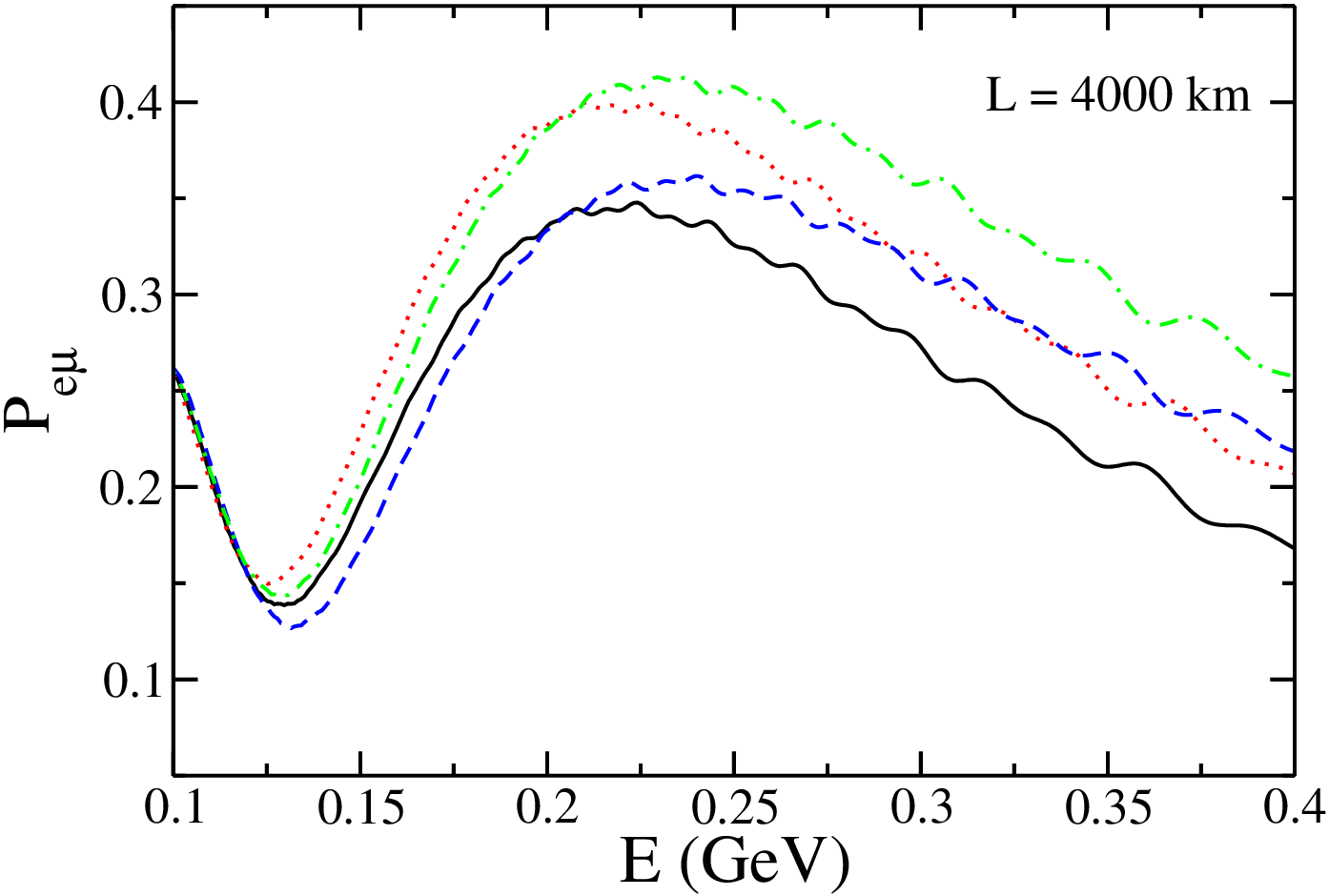}
\caption{[color online] The same as Fig.~\protect\ref{fixedl1} except with a baseline of 4000 km.}
\label{fixedl2}
\end{figure}
\begin{figure}[ht]
\includegraphics*[width=3in]{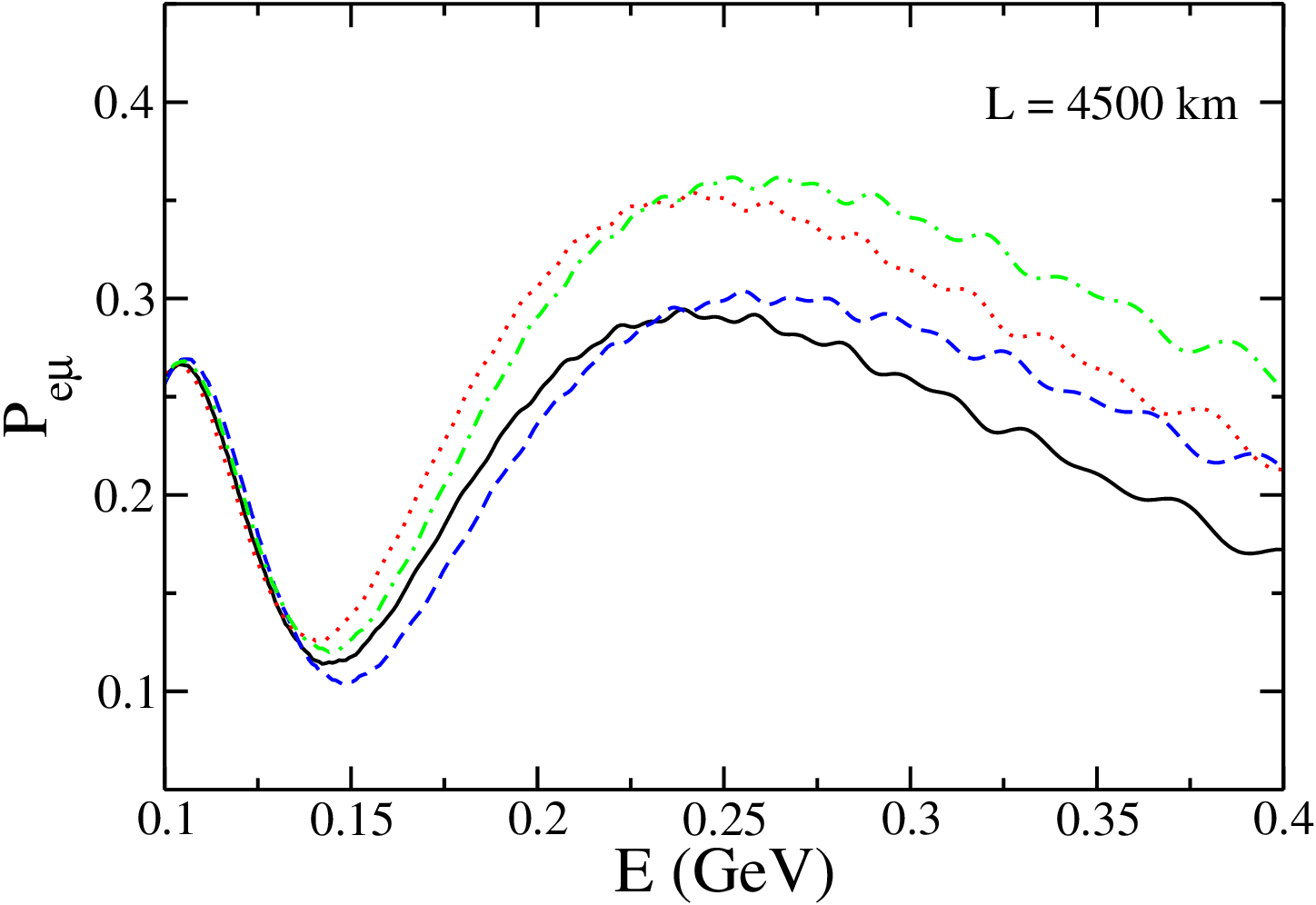}
\caption{[color online] The same as Fig.~\protect\ref{fixedl1} except with a baseline of 4500 km.}
\label{fixedl3}
\end{figure}
In Figs.~\ref{fixedl1}--\ref{fixedl3}, we plot $\mathcal{P}_{e \mu}$ for three very long baselines.
A horizontal shift (in $L/E$) of the location of the oscillation peaks is difficult to discern; however,  there is a considerable relative vertical shift in the height of the curves.  As this vertical shift is of import, we also plot the case in which CP is conserved but with negative mixing angle $\theta_{13}=-0.09$.  In Ref.~\cite{th13}, it was shown that a negative 
mixing  angle will shift the oscillation probability vertically at these baselines and energies.  Taken as a whole, there is no real signature amongst these curves which can charaterize both the level of CP violation and the sign of $\theta_{13}$.
However, if one assumed knowledge of $\theta_{13}$, then there is a clear separation between the CPC and CPV curves.
As an example, we consider the CPC and CPV curves for both having $\theta_{13} = 0.09$, the solid (black) curve
and the dotted (red) curve at the baseline of 4000 km. The percent difference between the peaks of these curves is 14\%. 
A similar relationship holds for the curves for $\theta_{13}=-0.09$, the dashed (blue) and the dash-dot (green) curves.
Thus, should $\theta_{13}$ become known and not be too small, this becomes a  possible experiment for searching for CP violation.  

\section{Conclusion}
We examined a means of detecting CP violation in a single neutrino oscillation channel. Such methods would 
require a non-zero value of $\theta_{13}$ as well as a precise knowledge of the small mass-squared difference. 
The CP phase was found to have the most profound impact for very long baselines at low energies, i.e., near the 
first peak of the
oscillatory region of the smaller mass-squared difference.  By examining a single oscillation channel, one 
avoids the possibility of a fake signal of CP violation attributable to matter effects.  Additionally, for the 
energies and baselines under consideration, mass hierarchy does not enter at a significant level.  It was shown 
that CP violation can be characterized by a shift (in $L/E$) of the peak of the appearance channel relative to CP 
conservation, both in vacuum and with MSW matter effects. In fact, matter effects enhance the shift at a 
fixed energy, and the effect is most pronounced for a varying baseline experiment rather than a fixed very long baseline. The best source for examining this effect is in atmospheric data, as fixing $E$ and varying $L$ is not so practical.  For this accurate knowledge of the absolute neutrino flux is not necessary; however, neutrinos must be distinguishable from antineutrinos.  
Unfortunately, the matter-induced energy dependence of the parameters obscures the shift (in $L/E$) of the peak at fixed baselines. However, the height of such peaks exhibits significant dependence on $\delta$ so that the level of CP violation could be resolved if $\theta_{13}$ were known. Finally, the proposed measurements likely could not resolve the $\theta_{13}$--$\delta$ degeneracy.

\acknowledgments{D.~C.~L.~thanks the Institute for Nuclear Theory at the University of Washington for its hospitality during 
this work. This work  was supported, 
in part, by U S Department of Energy Grant DE-FG02-96ER40975.

\bibliography{one_channel}
\end{document}